\documentclass[12pt]{article}

\begin{document}

\title{The Probability Distribution to Leptons and Quarks}
\author{G. Quznetsov \\
quznets@yahoo.com}
\date{April 14, 1999}
\maketitle

\begin{abstract}
The tracelike probability is expressed by the leptons and quarks
Hamiltonians.
\end{abstract}

I use the following notations \cite{LQH}:

\[
I=\left[ 
\begin{array}{cc}
1 & 0 \\ 
0 & 1
\end{array}
\right] \mbox{, }o=\left[ 
\begin{array}{cc}
0 & 0 \\ 
0 & 0
\end{array}
\right] 
\]

and the Pauli matrices:

\[
\sigma _x=\left( 
\begin{array}{cc}
0 & 1 \\ 
1 & 0
\end{array}
\right) ,\sigma _y=\left( 
\begin{array}{cc}
0 & -i \\ 
i & 0
\end{array}
\right) ,\sigma _z=\left( 
\begin{array}{cc}
1 & 0 \\ 
0 & -1
\end{array}
\right) \mbox{;} 
\]

three chromatic Clifford's pentads:

the red pentad $\zeta $:

\[
\zeta ^x=\left[ 
\begin{array}{cc}
\sigma _x & o \\ 
o & -\sigma _x
\end{array}
\right] ,\zeta ^y=\left[ 
\begin{array}{cc}
\sigma _y & o \\ 
o & \sigma _y
\end{array}
\right] ,\zeta ^z=\left[ 
\begin{array}{cc}
-\sigma _z & o \\ 
o & -\sigma _z
\end{array}
\right] , 
\]

\[
\gamma _\zeta ^0=\left[ 
\begin{array}{cc}
o & -\sigma _x \\ 
-\sigma _x & o
\end{array}
\right] \mbox{, }\zeta ^4=-i\cdot \left[ 
\begin{array}{cc}
o & \sigma _x \\ 
-\sigma _x & o
\end{array}
\right] ; 
\]

the green pentad $\eta $:

\[
\eta ^x=\left[ 
\begin{array}{cc}
-\sigma _x & o \\ 
o & -\sigma _x
\end{array}
\right] ,\eta ^y=\left[ 
\begin{array}{cc}
\sigma _y & o \\ 
o & -\sigma _y
\end{array}
\right] ,\eta ^z=\left[ 
\begin{array}{cc}
\sigma _z & o \\ 
o & \sigma _z
\end{array}
\right] , 
\]

\[
\gamma _\eta ^0=\left[ 
\begin{array}{cc}
o & -\sigma _y \\ 
-\sigma _y & o
\end{array}
\right] \mbox{, }\eta ^4=i\cdot \left[ 
\begin{array}{cc}
o & \sigma _y \\ 
-\sigma _y & o
\end{array}
\right] ; 
\]

the blue pentad $\theta $:

\[
\theta ^x=\left[ 
\begin{array}{cc}
\sigma _x & o \\ 
o & \sigma _x
\end{array}
\right] ,\theta ^y=\left[ 
\begin{array}{cc}
-\sigma _y & o \\ 
o & -\sigma _y
\end{array}
\right] ,\theta ^z=\left[ 
\begin{array}{cc}
\sigma _z & o \\ 
o & -\sigma _z
\end{array}
\right] , 
\]

\[
\gamma _\theta ^0=\left[ 
\begin{array}{cc}
o & -\sigma _z \\ 
-\sigma _z & o
\end{array}
\right] ,\theta ^4=-i\cdot \left[ 
\begin{array}{cc}
o & \sigma _z \\ 
-\sigma _z & o
\end{array}
\right] \mbox{;} 
\]

the light pentad $\beta $:

\[
\beta ^x=\left[ 
\begin{array}{cc}
\sigma _x & o \\ 
o & -\sigma _x
\end{array}
\right] ,\beta ^y=\left[ 
\begin{array}{cc}
\sigma _y & o \\ 
o & -\sigma _y
\end{array}
\right] ,\beta ^z=\left[ 
\begin{array}{cc}
\sigma _z & o \\ 
o & -\sigma _z
\end{array}
\right] , 
\]

\[
\gamma ^0=\left[ 
\begin{array}{cc}
o & I \\ 
I & o
\end{array}
\right] ,\beta ^4=i\cdot \left[ 
\begin{array}{cc}
o & I \\ 
-I & o
\end{array}
\right] \mbox{.} 
\]

Hence:

\[
\begin{array}{c}
\beta ^x=0.5\cdot \left( \beta ^x+\zeta ^x+\eta ^x+\theta ^x\right) \mbox{,}
\\ 
\beta ^y=0.5\cdot \left( \beta ^y+\zeta ^y+\eta ^y+\theta ^y\right) \mbox{,}
\\ 
\beta ^z=0.5\cdot \left( \beta ^z+\zeta ^z+\eta ^z+\theta ^z\right) .
\end{array}
\]

Let

\[
\left\langle \rho ,j_x,j_y,j_z\right\rangle 
\]

be a probability current vector \cite{PRS} and $\Psi $ be any complex
4-spinor \cite{LQH}:

\[
\Psi =\left| \Psi \right| \cdot \left[ 
\begin{array}{c}
\exp \left( i\cdot g\right) \cdot \cos \left( b\right) \cdot \cos \left(
a\right) \\ 
\exp \left( i\cdot d\right) \cdot \sin \left( b\right) \cdot \cos \left(
a\right) \\ 
\exp \left( i\cdot f\right) \cdot \cos \left( v\right) \cdot \sin \left(
a\right) \\ 
\exp \left( i\cdot q\right) \cdot \sin \left( v\right) \cdot \sin \left(
a\right)
\end{array}
\right] \mbox{.} 
\]

In this case the following system of equations

\[
\left\{ 
\begin{array}{c}
\Psi ^{\dagger }\cdot \Psi =\rho \mbox{,} \\ 
\Psi ^{\dagger }\cdot \beta ^x\cdot \Psi =j_x\mbox{,} \\ 
\Psi ^{\dagger }\cdot \beta ^y\cdot \Psi =j_y\mbox{,} \\ 
\Psi ^{\dagger }\cdot \beta ^z\cdot \Psi =j_z
\end{array}
\right| 
\]

has got the following type:

\[
\left\{ 
\begin{array}{c}
\Psi ^{\dagger }\cdot \Psi =\rho \mbox{,} \\ 
\left| \Psi \right| ^2\cdot \left( 
\begin{array}{c}
\cos ^2\left( a\right) \cdot \sin \left( 2\cdot b\right) \cdot \cos \left(
d-g\right) - \\ 
-\sin ^2\left( a\right) \cdot \sin \left( 2\cdot v\right) \cdot \cos \left(
q-f\right)
\end{array}
\right) =j_x \\ 
\left| \Psi \right| ^2\cdot \left( 
\begin{array}{c}
\cos ^2\left( a\right) \cdot \sin \left( 2\cdot b\right) \cdot \sin \left(
d-g\right) - \\ 
-\sin ^2\left( a\right) \cdot \sin \left( 2\cdot v\right) \cdot \sin \left(
q-f\right)
\end{array}
\right) =j_y \\ 
\left| \Psi \right| ^2\cdot \left( \cos ^2\left( a\right) \cdot \cos \left(
2\cdot b\right) -\sin ^2\left( a\right) \cdot \cos \left( 2\cdot v\right)
\right) =j_z
\end{array}
\right| \mbox{.} 
\]

Hence for every probability current vector: the spinor $\Psi $, obeyed to
this system, exists.

The operator $\widehat{U}\left( t,\triangle t\right) $, which acts in the
set of these spinors, is denoted as the evolution operator for the spinor $%
\Psi \left( t,\overrightarrow{x}\right) $, if:

\[
\Psi \left( t+\triangle t,\overrightarrow{x}\right) =\widehat{U}\left(
t,\triangle t\right) \Psi \left( t,\overrightarrow{x}\right) \mbox{.} 
\]

$\widehat{U}\left( t,\triangle t\right) $ is a linear operator.

The set of the spinors, for which $\widehat{U}\left( t,\triangle t\right) $
is the evolution operator, is denoted as the operator $\widehat{U}\left(
t,\triangle t\right) $ space.

The operator space is the linear space.

Let for an infinitesimal $\triangle t$:

\[
\widehat{U}\left( t,\triangle t\right) =1+\triangle t\cdot i\cdot \widehat{H}%
\left( t\right) \mbox{.} 
\]

Hence for an elements of the operator $\widehat{U}\left( t,\triangle
t\right) $ space:

\[
i\cdot \widehat{H}=\partial _t\mbox{.} 
\]

Since the functions $\rho $, $j_x$, $j_y$, $j_z$ fulfill to the continuity
equation \cite{PRS}:

\[
\partial _t\rho +\partial _xj_x+\partial _yj_y+\partial _zj_z=0 
\]

then:

\[
\left( \left( \partial _t\Psi ^{\dagger }\right) +\left( \partial _x\Psi
^{\dagger }\right) \cdot \beta _x+\left( \partial _y\Psi ^{\dagger }\right)
\cdot \beta _y+\left( \partial _z\Psi ^{\dagger }\right) \cdot \beta
_z\right) \cdot \Psi = 
\]

\[
=-\Psi ^{\dagger }\cdot \left( \left( \partial _t+\beta _x\cdot \partial
_x+\beta _y\cdot \partial _y+\beta _z\cdot \partial _z\right) \Psi \right) %
\mbox{.} 
\]

Let:

\[
\widehat{Q}=\left( i\cdot \widehat{H}+\beta _x\cdot \partial _x+\beta
_y\cdot \partial _y+\beta _z\cdot \partial _z\right) \mbox{.} 
\]

Hence:

\[
\Psi ^{\dagger }\cdot \widehat{Q}^{\dagger }\cdot \Psi =-\Psi ^{\dagger
}\cdot \widehat{Q}\cdot \Psi \mbox{.} 
\]

Therefore $i\cdot \widehat{Q}$ is the Hermitean operator.

Therefore:

\[
\widehat{H}=\beta _x\cdot \left( i\cdot \partial _x\right) +\beta _y\cdot
\left( i\cdot \partial _y\right) +\beta _z\cdot \left( i\cdot \partial
_z\right) -i\cdot \widehat{Q}\mbox{.} 
\]

Let

\[
-i\cdot \widehat{Q}= 
\]

\[
\left[ 
\begin{array}{cccc}
\varphi _{1,1} & \varphi _{1,2}+i\cdot \varpi _{1,2} & \varphi _{1,3}+i\cdot
\varpi _{1,3} & \varphi _{1,4}+i\cdot \varpi _{1,4} \\ 
\varphi _{1,2}-i\cdot \varpi _{1,2} & \varphi _{2,2} & \varphi _{2,3}+i\cdot
\varpi _{2,3} & \varphi _{2,4}+i\cdot \varpi _{2,4} \\ 
\varphi _{1,3}-i\cdot \varpi _{1,3} & \varphi _{2,3}-i\cdot \varpi _{2,3} & 
\varphi _{3,3} & \varphi _{3,4}+i\cdot \varpi _{3,4} \\ 
\varphi _{1,4}-i\cdot \varpi _{1,4} & \varphi _{2,4}-i\cdot \varpi _{2,4} & 
\varphi _{3,4}-i\cdot \varpi _{3,4} & \varphi _{4,4}
\end{array}
\right] \mbox{,} 
\]

here all $\varphi _{i,j}$ and $\varpi _{i,j}$ are a real functions on $%
R^{3+1}$.

Let:

\[
\left\{ 
\begin{array}{c}
B_0-B_z=\varphi _{3,3} \\ 
B_0+B_z=\varphi _{4,4}
\end{array}
\right| \mbox{,} 
\]

\[
\begin{array}{c}
B_x=\varphi _{3,4}\mbox{,} \\ 
B_y=\varpi _{3,4}\mbox{,}
\end{array}
\]

\[
\left\{ 
\begin{array}{c}
G+W_0=\varphi _{1,1}-\varphi _{4,4} \\ 
G-W_0=\varphi _{2,2}-\varphi _{3,3}
\end{array}
\right| \mbox{,} 
\]

\[
\begin{array}{c}
W_1=\varphi _{1,2}-\varphi _{3,4}\mbox{,} \\ 
W_2=-\varpi _{1,2}+\varpi _{3,4}\mbox{,}
\end{array}
\]

\[
\left\{ 
\begin{array}{c}
-a_\theta +a_\beta =\varphi _{1,3} \\ 
a_\theta +a_\beta =\varphi _{2,4}
\end{array}
\right| 
\]

\[
\left\{ 
\begin{array}{c}
b_\beta -b_\theta =\varpi _{1,3} \\ 
b_\beta +b_\theta =\varpi _{2,4}
\end{array}
\right| 
\]

\[
\cos \left( \alpha _\theta \right) =\frac{a_\theta }{\sqrt{a_\theta
^2+b_\theta ^2}}\mbox{, }\sin \left( \alpha _\theta \right) =\frac{b_\theta 
}{\sqrt{a_\theta ^2+b_\theta ^2}} 
\]

\[
\cos \left( \alpha _\beta \right) =\frac{a_\beta }{\sqrt{a_\beta ^2+b_\beta
^2}}\mbox{, }\sin \left( \alpha _\beta \right) =\frac{b_\beta }{\sqrt{%
a_\beta ^2+b_\beta ^2}} 
\]

\[
m_\theta =2\cdot \sqrt{a_\theta ^2+b_\theta ^2}\mbox{, }m_\beta =2\cdot 
\sqrt{a_\beta ^2+b_\beta ^2} 
\]

\[
\begin{array}{c}
\gamma _\theta =\left( \cos \left( \alpha _\theta \right) \cdot \gamma
_\theta ^0+\sin \left( \alpha _\theta \right) \cdot \theta ^4\right) \\ 
\gamma _\beta =\left( \cos \left( \alpha _\beta \right) \cdot \gamma _\beta
^0+\sin \left( \alpha _\beta \right) \cdot \beta ^4\right)
\end{array}
\]

(here: for $\mu \in \left\{ x,y,z\right\} $:

\[
\begin{array}{c}
\gamma _\theta \cdot \theta ^\mu =-\theta ^\mu \cdot \gamma _\theta 
\mbox{
and} \\ 
\gamma _\theta \cdot \gamma _\theta =1_4\mbox{;}
\end{array}
\]

\[
\begin{array}{c}
\gamma _\beta \cdot \beta ^\mu =-\beta ^\mu \cdot \gamma _\beta \mbox{ and}
\\ 
\gamma _\beta \cdot \gamma _\beta =1_4\mbox{;}
\end{array}
\]

see \cite{H})

\[
\left\{ 
\begin{array}{c}
-a_\zeta +b_\eta =\varphi _{1,4} \\ 
-a_\zeta -b_\eta =\varphi _{2,3}
\end{array}
\right| 
\]

\[
\left\{ 
\begin{array}{c}
a_\eta -b_\zeta =\varpi _{1,4} \\ 
-a_\eta -b_\zeta =\varpi _{2,3}
\end{array}
\right| 
\]

\[
\cos \left( \alpha _\zeta \right) =\frac{a_\zeta }{\sqrt{a_\zeta ^2+b_\zeta
^2}}\mbox{, }\sin \left( \alpha _\zeta \right) =\frac{b_\zeta }{\sqrt{%
a_\zeta ^2+b_\zeta ^2}} 
\]

\[
\cos \left( \alpha _\eta \right) =\frac{a_\eta }{\sqrt{a_\eta ^2+b_\eta ^2}}%
\mbox{, }\sin \left( \alpha _\eta \right) =\frac{b_\eta }{\sqrt{a_\eta
^2+b_\eta ^2}} 
\]

\[
m_\zeta =2\cdot \sqrt{a_\zeta ^2+b_\zeta ^2}\mbox{, }m_\eta =2\cdot \sqrt{%
a_\eta ^2+b_\eta ^2} 
\]

\[
\begin{array}{c}
\gamma _\zeta =\left( \cos \left( \alpha _\zeta \right) \cdot \gamma _\zeta
^0+\sin \left( \alpha _\zeta \right) \cdot \zeta ^4\right) \\ 
\gamma _\eta =\left( \cos \left( \alpha _\eta \right) \cdot \gamma _\eta
^0+\sin \left( \alpha _\eta \right) \cdot \eta ^4\right)
\end{array}
\]

(here: for $\mu \in \left\{ x,y,z\right\} $:

\[
\begin{array}{c}
\gamma _\zeta \cdot \zeta ^\mu =-\zeta ^\mu \cdot \gamma _\zeta \mbox{ and}
\\ 
\gamma _\zeta \cdot \gamma _\zeta =1_4\mbox{;}
\end{array}
\]

\[
\begin{array}{c}
\gamma _\eta \cdot \eta ^\mu =-\eta ^\mu \cdot \gamma _\eta \mbox{ and} \\ 
\gamma _\eta \cdot \gamma _\eta =1_4\mbox{;}
\end{array}
\]

)

In this case:

\[
\widehat{H}= 
\]

\[
=0.5\cdot \left( 
\begin{array}{c}
i\cdot \left( \beta ^x\cdot \left( \partial _x-i\cdot B_x\right) +\beta
^y\cdot \left( \partial _y-i\cdot B_y\right) +\beta ^x\cdot \left( \partial
_z-i\cdot B_z\right) \right) + \\ 
+m_\beta \cdot \gamma _\beta
\end{array}
\right) + 
\]

\[
+0.5\cdot \left( 
\begin{array}{c}
i\cdot \left( \zeta ^x\cdot \left( \partial _x-i\cdot B_x\right) +\zeta
^y\cdot \left( \partial _y-i\cdot B_y\right) +\zeta ^x\cdot \left( \partial
_z-i\cdot B_z\right) \right) + \\ 
+m_\zeta \cdot \gamma _\zeta
\end{array}
\right) + 
\]

\[
+0.5\cdot \left( 
\begin{array}{c}
i\cdot \left( \theta ^x\cdot \left( \partial _x-i\cdot B_x\right) +\theta
^y\cdot \left( \partial _y-i\cdot B_y\right) +\theta ^x\cdot \left( \partial
_z-i\cdot B_z\right) \right) + \\ 
+m_\theta \cdot \gamma _\theta
\end{array}
\right) + 
\]

\[
+0.5\cdot \left( 
\begin{array}{c}
i\cdot \left( \eta ^x\cdot \left( \partial _x-i\cdot B_x\right) +\eta
^y\cdot \left( \partial _y-i\cdot B_y\right) +\eta ^x\cdot \left( \partial
_z-i\cdot B_z\right) \right) + \\ 
+m_\eta \cdot \gamma _\eta
\end{array}
\right) + 
\]

\[
+\left[ 
\begin{array}{cccc}
W_0 & W_1-i\cdot W_2 & 0 & 0 \\ 
W_1+i\cdot W_2 & -W_0 & 0 & 0 \\ 
0 & 0 & 0 & 0 \\ 
0 & 0 & 0 & 0
\end{array}
\right] + 
\]

\[
+\left[ 
\begin{array}{cccc}
G & 0 & 0 & 0 \\ 
0 & G & 0 & 0 \\ 
0 & 0 & 0 & 0 \\ 
0 & 0 & 0 & 0
\end{array}
\right] + 
\]

\[
+\left[ 
\begin{array}{cccc}
B_0 & 0 & 0 & 0 \\ 
0 & B_0 & 0 & 0 \\ 
0 & 0 & B_0 & 0 \\ 
0 & 0 & 0 & B_0
\end{array}
\right] \mbox{.} 
\]

\end{document}